\RequirePackage{amsmath}
\documentclass[runningheads]{llncs}
\usepackage{graphicx}
\usepackage{amsfonts}
\usepackage{amssymb}
\usepackage{amsmath}
\usepackage{enumerate}

\begin{document}
\title{On the linear complexity of feedforward clock-controlled sequence}
%
%\titlerunning{Abbreviated paper title}
% If the paper title is too long for the running head, you can set
% an abbreviated paper title here
%
\author{Yangpan Zhang\inst{1} \and Maozhi Xu\inst{1}}
%
%\authorrunning{F. Author et al.}
% First names are abbreviated in the running head.
% If there are more than two authors, 'et al.' is used.
%
\institute{School of Mathematical Sciences, Peking University, Beijing 100871, China\\ \email{zyp94@pku.edu.cn}}
%\institute{Princeton University, Princeton NJ 08544, USA \and
%Springer Heidelberg, Tiergartenstr. 17, 69121 Heidelberg, Germany
%\email{lncs@springer.com}\\
%\url{http://www.springer.com/gp/computer-science/lncs} \and
%ABC Institute, Rupert-Karls-University Heidelberg, Heidelberg, Germany\\
%\email{\{abc,lncs\}@uni-heidelberg.de}}
%
\maketitle              % typeset the header of the contribution
\pagestyle{plain}
\begin{abstract}

  As a research field of stream ciphers, the pursuit of a balance of security and practicality is the focus. The conditions for security
  usually have to satisfy at least high period and high linear complexity. Because the feedforward clock-controlled structure can provide
  quite a high period and utility, many sequence ciphers are constructed based on this structure. However, the past study of its linear
  complexity only works when the controlled sequence is an m-sequence.  Using the theory of matrix over the ring and block matrix in this
  paper, we construct a more helpful method. It can estimate the lower bound of the linear complexity of the feedforward clock-controlled
  sequence. Even the controlled sequence has great linear complexity.

\end{abstract}

\section{Introduction}

A clock-controlled structure is a structure that uses one sequence generator as a clock to control another sequence generator (or control
itself) to generate a new sequence. The sequences generated by this structure have a large linear complexity and are widely used in stream
cipher design.

The first proposal of the clock-controlled structure dates back to 1980 when Jennings\cite{Jennings1980A} and Kjeldsen\cite{KjeldsenK1980Some}
proposed a similar structure, respectively. In 1984, T. Beth and F. C. Piper\cite{beth1984stop} first introduced the concept of
"clock-controlled."

The subsequent studies\cite{gollmann1989clock} divided the clock-controlled structure into two categories, i.e., feedforward and feedback
clock-controlled. The basic feedforward clock-controlled structure refers to using a regular sequence generator to control the clock of another
sequence generator. For the feedback clock-controlled structure,  it uses the output of the pseudo-random sequence generator to clock-control
itself. In practice, the feedback structure makes it challenging to analyze the security from the theory, so most of the clock-controlled
sequences are of feedforward structure.

The feedforward clock-controlled structure has a mathematically more apparent structure and better theoretical analysis results for its
periodic and statistical properties\cite{kholosha2004investigations}. However, the study of linear complexity is not as clear.

The upper bound on the linear complexity is $nq$\cite{XiangangL1990}, where the order $n$ is the linear complexity of
the controlled sequence, and $q$ is the period of the control sequence. However, the conditions for the linear complexity to reach the upper
bound are pretty demanding.

By analyzing irreducible polynomials over a finite field, assuming that the controlled sequence is an $m$-sequence, Li finds a sufficient
condition for the linear complexity to reach an upper bound\cite{XiangangL1990}. In contrast, Golic J.D analyzes it from
a probabilistic point of view in 1988\cite{golic1988linear}. The probability of the linear complexity reaching the upper bound tends to 1 as
$n$ grows. When the controlled sequence is an $m$-sequence of order $n$, the step sum $M$ is less than $2^n$.

The above studies were published around 1990. However, in the last years of the 20th century, stream cryptanalysis tools such as linear
analysis \cite{matsui1993linear}\cite{coppersmith2002cryptanalysis}, correlation analysis
\cite{golic1996correlation}\cite{meier1990correlation}, and algebraic attacks \cite{courtois2003algebraic} were widely researched and
developed. The discovery of these analysis tools has made the traditional sequence cryptosystem based on LFSR design less secure. People
gradually abandoned the design approach using LFSRs as linear drivers and shifted to nonlinear design schemes. In this way, the above-mentioned
linear complexity study of clock-controlled sequences based on $m$-sequences was rendered useless.

Furthermore, when the controlled sequence is nonlinear, its minimal polynomials are often reducible and irregular. Even the linear complexity
is unknown.  Therefore, in practical analysis, people tend to use less rigorous experimental analysis methods. That is, analyze the actual
linear complexity in the degenerate case with shortened register. Then the nondegenerate case is reasonably guessed by the relationship between
register length and linear complexity. Such as the LILI-128 algorithm \cite{dawson2000lili}.

In this paper, we make a new method to estimate the lower bound of linear complexity of a feedforward clock-controlled sequence. This new
method can estimate better when the clock-controlled sequence is under a nonlinear driver. Unlike the current result, this paper does not
analyze the polynomial reducibility. However, it estimates the lower bound of the matrix rank of the sequence-generating circulant matrix after
a proper transformation. Our approach method gives a better bound on the linear complexity of the feedforward clock-controlled sequence. Unlike
the current results in the papers \cite{XiangangL1990}\cite{golic1988linear}\cite{XiangangL1991}, this method
does not require the controlled sequence to be an $m$-sequence. It is, therefore, suitable for feedforward clock-controlled sequences in a
general sense.

The article is structured as follows. Section 2 will give the basic concepts in the study and some mathematical tools for the study of block
matrices. With the help of these tools, we give in Section 3 an estimation method for the lower bound of the linear complexity of the
feedforward clock-controlled sequence. Section 4 proposes its improved algorithm LIFI-128 based on the LILI-128 algorithm and estimates its
linear complexity very well. A summary of the whole paper is given in Section 5.

\section{Pre-requisite knowledge}
\subsection{feedforward clock-controlled sequence}
The paper \cite{gollmann1989clock} is a good review of clock-controlled shift registers, after which the definition of a basic feedforward
clock-controlled sequence generator can be given as follows.

\begin{definition}[Basic clock-controlled sequence generator]

\begin{itemize}
\item[\bf Input:] a Control Sequence Generator $A$ with period $T_1$; a Controlled Sequence Generator $B$ with period $T_2$; a step map
    $f_L:output_A\rightarrow Z_{T_2}$. where $output_A$ represents the set of possible states of the output of generator $A$ at any moment.
\item[\bf Key:] the initial states of the two sequence generators $A$ and $B$.
\item[\bf Process:] Denote the initial state moment as $t=0$. For $t=1,2,\cdots$, complete the following actions step by step.
	\subitem{1} Run sequence generator $A$ for one time, after which the current output state of sequence generator $A$ is recorded as $a_t$,
and $f_L\left(a_t\right)$ is calculated.
    \subitem{2} Run the sequence generator $B$ for a total of $f_L\left(a_t\right)$ times, after which the state $b_{\sigma_t}$ of the output
    of $B$ is set to the output state $c_t=b_{\sigma_t}$ of the clock-controlled sequence generator at moment $t$. where $b_i$ is the output
    state of generator $B$ after continuous running $i$ times since the initial state,
    $\sigma_t=\sum_{i=1}^{t}\left(f_L\left(a_i\right)\right).$
\item[\bf Output:] clock-controlled sequence $\{c_t\}_{t=1}^\infty$.
\end{itemize}
\end{definition}

In the above definition, we call the sequence generated by $A$ under the action of $f_L$ a Control Sequence and the sequence generated by $B$
under the control of a regular clock a Controlled Sequence.

This definition can also be reduced to a binary pseudo-random sampling sequence as follows.

\begin{definition}[Binary pseudo-random sampling sequence]
\begin{itemize}
    \item[\bf Input:] given a binary periodic sequence $\{b_t\}=\left(b_0,b_1,\cdots,\right)$, where $b_i\in F_2$; given a pseudo-random
        sampling subscript sequence $\{\sigma_t\}=\left(\sigma_1,\sigma_2,\cdots\right)$, where $\sigma_i\in N$.
    \item[\bf Output:] a new set of binary sequences $\{c_t\}=\left(b_{\sigma_1},b_{\sigma_2},\cdots\right)$. Call it a pseudo-random
        sampling sequence.
\end{itemize}
\end{definition}

For the period of the clock-controlled sequence, the following result is obtained.

\begin{theorem}\cite{blakley1981necessary}
Denote $S=\sum_{i=1}^{T_1}\left(f_L\left(a_i\right)\right)$, i.e., $S=\sigma_{T_1}$. When $gcd\left(S,T_2\right)=1$, i.e., when the integer $S$
is coprime with the period $T_2$. The minimum positive period of the clock-controlled sequence $\{c_t\}_{t=1}^\infty$ is $T_3=T_1T_2$, which
reaches a maximum period.
\end{theorem}

For clock-controlled sequence algorithms, the maximum period is always preferred in practical applications. Therefore, all the sequence models
for clock control that appear below in this paper are chosen to reach the maximum period.

\subsection{Linear complexity and circulant matrix}

In recent years, the LFSR structure is no longer directly used to construct stream cipher regimes. However, the linear complexity also measures
the resistance of a sequence to many linear-based attacks.  Therefore, linear complexity is still a very important metric in measuring stream
cipher security.

An equivalent definition of linear complexity is given below after the definition of circulant matrix.

\begin{definition}
On a field $\mathbb{K}$, a matrix of the following shape is called a $n\times n$ $r$-circulant matrix. where $r \in K$.
\begin{equation}
\left(\begin{matrix}a_1&a_2&a_3&\cdots&a_{n-1}&a_n\\ra_n&a_1&a_2&\cdots&a_{n-2}&a_{n-1}\\{ra}_{n-1}&{ra}_n&a_1&\cdots&a_{n-3}&a_{n-2}\\\vdots&\vdots&\vdots&\vdots&\vdots&\vdots\\{ra}_2&ra_3&{ra}_4&\cdots&{ra}_n&a_1\\\end{matrix}\right)_{n\times
n}
\end{equation}

For convenience, it can be generally shortened to $Cir^{r}_n\left(a_1,a_2,\cdots,a_n\right)$. Specially, if $r=1$, we call it circulant matrix.
\end{definition}

For a purely periodic sequence $A=\left(a_1,a_2,\cdots\right)$ of period $n$ over a field $\mathbb{K}$. Denote
$Cir^{1}_n\left(a_1,a_2,\cdots,a_n\right)$ by $M_{cir}(A)$.

\begin{theorem}\cite{schaub1990linear}
$A$ is a purely periodic sequence on a field $\mathbb{K}$ with period $n$. Then, for $M_{cir}\left(A\right)$, there is such a property. That
is, the rank of $M_{cir}\left(A\right)$ is equal to the linear complexity $L\left(A\right)$ of the sequence $A$ over the field $\mathbb{K}$.
\end{theorem}

When the sequence $B=\left(b_1,b_2,\cdots\right)$, is regular sampled from the sequence $A=\left(a_1,a_2,\cdots\right)$, with a period of $l$.
That is, for any $i\geq 1$, we have $b_i=a_{s+l\cdot\left(i-1\right)}$, where $b_1=a_s$ is called the starting sampling point. It can be
denoted briefly as $B=A\left(s,l\right)$. If $A$ is a sequence of period $n$ and satisfies $gcd\left(l,n\right)=1$, then the following
corollary can be obtained using Theorem 2.

\begin{corollary}
Assume $A$ is a purely periodic sequence over a field $\mathbb{K}$ with period $n$. And the sequence $B=A\left(s,l\right)$ is a sequence of
regular samples of the sequence $A$. If $gcd\left(l,n\right)=1$, then: (1) the period of sequence $B$ is $n$; (2) $L(A)=L(B)$.
\end{corollary}

The proof of the corollary is simple; it only requires a proper primary rows and columns swap for $M_{cir}\left(A\right)$ to become
$M_{cir}\left(B\right)$.  Therefore, the two sequences have the same linear complexity.

For any $r$-circulant matrix over a number field $\mathbb{K}$, there is a very important theorem.

\begin{theorem}\cite{cline1974generalized}
Let $M=Cir_n^r\left(w_1,w_2,\cdots,w_n\right)$ be an $r$-circulant matrix over field $\mathbb{K}$.Denote the function
$w\left(x\right)=\sum_{i=0}^{n-1}w_{i+1}x^i$. If the set of all roots of the equation $x^n-r=0$ over field $\mathbb{K}$ can be written as
$\{\theta\xi^i\ |\ i=0,1,\cdots,\ n-1\}$, where $\theta^n=r$. Then the set of all characteristic roots of the matrix $M$ is $\{w(\theta\xi^i)\
|\ i=0,1,\cdots,\ n-1\}$.
\end{theorem}

\subsection{Block matrix and matrix over ring}
Let $\mathbb{K}$ be a field, denote the ring of all $mn\times mn$ matrices over field $\mathbb{K}$ by $M_{mn\times mn}(\mathbb{K})$. Mark
matrix ring $R$ as subring of $M_{n\times n}(\mathbb{K})$. Suppose a matrix $A$ belongs to $M_{m\times m}(R)$, then $A$ also belongs to
$M_{mn\times mn}(\mathbb{K})$. Let $[A]_{i,j}^{R}$ denote the $i,j$th block of $A$, $a_{i,j}$ denote the $i,j$th entry of $A$ when over
$M_{mn\times mn}(\mathbb{K})$. It's easy to see that

$$
    A =
        \left(\begin{matrix}
            a_{1,1} & a_{1,2} & \cdots & a_{1,mn} \\
            a_{2,1} & a_{2,2} & \cdots & a_{2,mn} \\
            \vdots  & \vdots  & \ddots  & \vdots \\
            a_{mn,1} & a_{mn,2} & \cdots & a_{mn,mn} \\
        \end{matrix}\right)
       =
        \left(\begin{matrix}
            A_{1,1} & A_{1,2} & \cdots & A_{1,m} \\
            A_{2,1} & A_{2,2} & \cdots & A_{2,m} \\
            \vdots  & \vdots  & \ddots  & \vdots \\
            A_{m,1} & A_{m,2} & \cdots & A_{m,m} \\
        \end{matrix}\right)
        = A^{R}
$$

The above sliced matrix $A$ is called the block matrix, In particular, when we discuss $A$ as a element of $M_{m\times m}(R)$, we use $A^R$ to
denote $A$, and the corner marks are used only for distinction.

For a general commutative ring $R$, Brown W C \cite{brown1993matrices} studied relevant properties about matrices over the ring $R$. Including
the determinant $det_{R}(A^{R})$, rank $rank_{R}(A^{R})$, modulus, diagonalization. Based on the definitions and results given in the book, we
got the following remarkable theorems.

\begin{theorem}
Let $A\in M_{m\times m}\left(R\right)$, where $R=\{\sum_{i=0}^{\infty}k_{i}S^{i}| k_{i}\in \mathbb{K}\}$ is a subalgebra of $M_{n\times n}\left(\mathbb{K}\right)$. In particular, the minimal polynomial
$f\left(x\right)=p^r\left(x\right)$ of $S\in M_{n\times n}\left(\mathbb{K}\right)$ is an power of an irreducible polynomial $p(x)$ over the
field $\mathbb{K}$. Thus,

$$
rank_{R}\left(A^{R}\right)=k\Rightarrow rank_{\mathbb{K}}\left(A\right)\geq kn, 0\leq k \leq m
$$
\end{theorem}

Clearly, when the commutative ring $R$ satisfies the conditions in the above theorem, $R$ is isomorphic to the residue class ring $H=
\mathbb{K}[x]/(p^{r}(x))$. This means that the equation $rank_{R}(A^{R}) = rank_{H}(A^{H})$ will hold automatically under isomorphism.

Denote another ring of residue classes $\overline{H} = \mathbb{K}[x]/(p(x))$, it's easy to see $\overline{H}$ is a field. At the same time,
there exists a surjective homomorphism mapping $\pi$ from $H$ to $\overline{H}$. The image of $A^{H}$ under the action of $\pi$ is written as
$\overline{A} \in M_{m\times m}(\overline{H})$. We have the following theorem.

\begin{theorem}
$$rank_{H}\left(A^H\right)=rank_{\overline{H}}\left(\overline{A}\right)$$
\end{theorem}

These two theorems provide theoretical support for our estimate of the lower bound on linear complexity. The proof procedure is complex and
unproductive for this paper. For logical reasons, the exact process of their proof is omitted.

\section{Linear complexity estimation model for feedforward clock-controlled sequences}

This section we will show you how to use the basic model of pseudo-random sampling. And transform sequences' circulant matrix. Finally estimate
the rank of block matrix.

Denote two period sequence $\{a_{i}\}_{\infty}$ and $\{b_{i}\}_{\infty}$, where $a_{i}\in \mathbb{N}$ and $b_{i}\in \mathbb{F}_{2}$. Denote
$\sum_{i = 1}^{k}a_{i}$ by $s_{k}$. By sampling $\{b_{i}\}_{\infty}$ with index sequence $\{s_{i}\}_{\infty}$, we get a new sequence
$C=\{c_{i}\}_{\infty}$, where $c_{i} = b_{s_{i}}$. We call $\{c_{i}\}_{\infty}$ a clock-controlled sequence generated by $\{a_{i}\}_{\infty}$
controlling $\{b_{i}\}_{\infty}$.

In general case, people prefer to use maximal period sequences as them have good statistical properties. So we always assume $s_{m}$ is coprime
with the period $n$ in follow discussion.

\subsection{Primary transformation of the circulant matrix $M_{cir}(C)$}
It's hard to direct calculate rank of $M_{cir}(C)$, so we do some row operations and column operations on $M_{cir}(C)$ and denote the matrix
after operations by $\overline{C}$:

\begin{equation*}
\bordermatrix{%
        &1        &2      & 3     & \cdots & mn         \cr
1       &c_{1}    &c_{2}  & c_{3} & \cdots & c_{mn}     \cr
2       &c_{mn}   &c_{1}  & c_{2} &        & c_{mn-1}   \cr
3       &c_{mn-1} &c_{mn} & c_{1} &        & c_{mn-2}   \cr
\vdots  &\vdots   &       &       & \ddots & \vdots     \cr
mn      &c_{2}    &c_{3}  & c_{4} & \cdots & c_{1}      \cr
}
\Rightarrow
\bordermatrix{%
        &I_{1}    &I_{2}  & I_{3}  & \cdots & I_{m}      \cr
I_{1}   &C_{1,1}  &C_{1,2}& C_{1,3}&        & C_{1,m}    \cr
I_{2}   &C_{2,1}  &C_{2,2}& C_{2,3}& \cdots & C_{2,m}    \cr
I_{3}   &C_{3,1}  &C_{3,2}& C_{3,3}&        & C_{3,m}    \cr
\vdots  &         &\vdots &        & \ddots & \vdots     \cr
I_{m}   &C_{m,1}  &C_{m,2}& C_{m,3}& \cdots & C_{m,m}    \cr
}
\end{equation*}

Where the index set $I_{i} = \{i,m+i,2m+i,\cdots,(n-1)m+i\}$, and the submatrix $C_{i,j}$ was construct by entries from $I_{i}$'s rows and
$I_{j}$'s columns of $M_{cir}(C)$. Assume $t = (j - i + 1) \mod mn$, then:
\begin{equation*}
C_{i,j} =
\bordermatrix{%
        &j              &m+j            & 2m+j      & \cdots & (n-1)m+j     \cr
i       &c_{t}          &c_{t+m}        & c_{t+2m}  &        & c_{t+(n-1)m} \cr
m+i     &c_{t+(n-1)m}   &c_{t}          & c_{t+m}   & \cdots & c_{t+(n-2)m} \cr
2m+i    &c_{t+(n-2)m}   &c_{t+(n-1)m}   & c_{t}     &        & c_{t+(n-3)m} \cr
\vdots  &               &\vdots         &           & \ddots & \vdots       \cr
(n-1)m+i&c_{t+m}        &c_{t+2m}       & c_{t+3m}  & \cdots & c_{t}        \cr
}
\end{equation*}

It's easy to show that $C_{i,j}$ was a circulant matrix, and for two submatrices $C_{i,j}$ and $C_{i',j'}$, $C_{i,j} = C_{i',j'}$ if and only
if $j-i = j'-i'$.

Consider subsequence $C^{t} = \{c_{t+m\cdot i}\}_{\infty}$, this sequence has a period of $n$. In fact, $C^{t}$ equals to $\{b_{s_{t} +
s_{m}\cdot i}\}_{\infty}$, it's a sampling sequence of $\{b_{i}\}_{\infty}$ with $s_{m}$ step length. Further, assume $v = (s_{m})^{-1} \mod n$
and $l_{t} = v(s_{t} - s_{1})$, $C^{t}$ equals to $C^{1}$ start from $l_t$th position.

Using the fact that $C_{i,j}$ is a circulant matrix, $C_{i,j}$ equals to $M_{cir}(C^{t})$. Thus, there is a formula:
\begin{equation}
C_{i,j} = M_{cir}(C^{t}) = M_{cir}(C^{1})\cdot D^{l_{t}}
\end{equation}
$D$ is a primitive circulant matrix with dimension $n$, as shown in follow:

$$
D =
\left(
\begin{matrix}
0 & 1 & 0 & 0 &        & 0\\
0 & 0 & 1 & 0 & \cdots & 0\\
0 & 0 & 0 & 1 &        & 0\\
  &\vdots&   &   &\ddots &\\
1 & 0 & 0 & 0 & \cdots   & 0\\
\end{matrix}\right).
$$

Turn back to $C_{i,j}$, if $j \geq i$, then $t = j - i + 1$, if $j < i$, then $t = j - i + 1 + mn$. So $$C_{i,j} = M_{cir}(C^{1})\cdot
D^{l_{t}} = M_{cir}(C^{1})\cdot D^{v(s_{t} - s_{1})}.$$

Denote $D^{v}$ by $T$, denote $M_{cir}(C^{1})\cdot T^{-s_{1}}$ by $\hat{C}$. Notice that $T^{n} = I$, we denote $s_{-i} = s_{m-i} - s_{m}$.
When $j \geq i$, $C_{i,j} = \hat{C}\cdot T^{s_{j-i+1}}$; when $j < i$, $C_{i,j} = \hat{C}\cdot T^{s_{j-i+1} + s_{m}n} = \hat{C}\cdot
T^{s_{j-i+1}} = \hat{C}\cdot T^{s_{m+j-i+1} - s_{m}} = \hat{C}\cdot T^{s_{m+j-i+1}}\cdot D^{-1}$. Different premise get same result.

Thus,
\begin{equation*}
\overline{C} =
\begin{pmatrix}
\hat{C}T^{s_{1}} & \hat{C}T^{s_{2}} & \hat{C}T^{s_{3}} &  & \hat{C}T^{s_{m} }\\
\hat{C}T^{s_{m} }D^{-1} & \hat{C}T^{s_{1}} & \hat{C}T^{s_{2} } & \cdots  &\hat{C}T^{s_{m-1}}\\
\hat{C}T^{s_{m-1} }D^{-1} & \hat{C}T^{s_{m} }D^{-1} & \hat{C}T^{s_{1}} & & \hat{C}T^{s_{m-2} }\\
&\vdots& &\ddots & \vdots\\
\hat{C}T^{s_{2}}D^{-1} & \hat{C}T^{s_{3} }D^{-1} & \hat{C}T^{s_{4} }D^{-1} &\cdots & \hat{C}T^{s_{1}}\\
\end{pmatrix}.
\end{equation*}

\subsection{Decomposition of the matrix over the ring}
In this part, some Lemmas are needed to decompose the matrix over the ring $R = <S> = \left\{\sum_{i=0}^{\infty}k_iS^i\middle| k_i\in
\mathbb{K},i=0,1,\cdots\right\}$.

Let $R = <S>=\{\sum_{i=0}^{\infty}k_{i}S^{i}|k_{i}\in \mathbb{K}, i=0,1,\cdots\}$, where $S$ is an element of $M_{n\times n}(\mathbb{K})$. It's
obvious that $R$ is a commutative(multiplication) subalgebra of $M_{n\times n}(\mathbb{K})$

Denote $S$'s minimal polynomial over $\mathbb{K}$ by $f(x)=\sum_{i=0}^{l}f_{i}x^{i}$, where $l \leq n$ and $f_{i}\in \mathbb{K}$($f_{l} = 1$).
Thus,
$$
<S> = \{\sum_{i=0}^{l-1}k_{i}S^{i}|k_{i}\in \mathbb{K}, i=0,1,\cdots,l-1\}.
$$
Given $U,V\in <S>$, where $U = \sum_{i=0}^{l-1}u_{i}S^{i}, V = \sum_{i=0}^{l-1}v_{i}S^{i}$. It's obvious that $U = V$ if and only if $u_{i} =
v_{i}, \forall i=0,1,\cdots,l-1$.

\begin{lemma}
Suppose that $S\in M_{n\times n}(\mathbb{K})$ is a matrix over the field $\mathbb{K}$, where the minimal polynomial of $S$ is $f(x)$. And the
unique factorization of $f(x)$ over field $K$ is $f\left(x\right)=\prod_{i=1}^{d}{p_i^{r_i}\left(x\right)}$, $p_{i}(x)$ is irreducible and
$p_{i}(x) \neq p_{j}(x)$ when $i\neq j$.

 Thus, there exists a non-singular matrix $P\in M_{n\times n}\left(\mathbb{K}\right)$, and matrices $S_{i}$ for $1\leq i \leq d$. Where the
 minimal polynomial of $S_{i}$ is $p^{r_{i}}_{i}(x)$. Such that:
\begin{equation*}
S=P^{-1}\cdot\left(\begin{matrix}S_1&&&\\&S_2&&\\&&\ddots&\\&&&S_d\\\end{matrix}\right)\cdot P
\end{equation*}
\end{lemma}

In the classical theory of linear algebra, this lemma can be easily proved by analyzing the invariant subspace of the linear transformation.

\begin{corollary}
For any $U=\sum_{i=0}^{l-1}u_i S^i\in R = <S>$, exist mapping $g(U) = P\cdot U\cdot P^{-1}$, from $R=<S>$ to $\overline{R} = <PSP^{-1}>$. And,
$$
g\left(U\right)=\left(\begin{matrix}\sum_{i=0}^{l-1}u_iS_1^i&&\\&\ddots&\\&&\sum_{i=0}^{l-1}u_iS_d^i\\\end{matrix}\right)
$$
\end{corollary}

Extend the mapping $g$ from $M_{n\times n}\left(\mathbb{K}\right)$ to $M_{m\times m}\left(M_{n\times n}\left(\mathbb{K}\right)\right)$. Define
a mapping $G$ on $M_{m\times m}\left(M_{n\times n}\left(\mathbb{K}\right)\right)$. for any element $T\in M_{m\times m}\left(M_{n\times
n}\left(\mathbb{K}\right)\right )$, $T$ can be written as block matrix $T^{M_{n\times n}(\mathbb{K})} = \left(T_{ij}\right)_{m\times m}$, where
$T_{ij} \in M_{n\times n}\left(\mathbb{K}\right)$. The mapping $G$ is defined as:
$$
G((T_{ij})_{m\times m}) = (g(T_{ij}))_{m\times m}.
$$

Obviously, $G$ is a self-isomorphism on $M_{m\times m}\left(M_{n\times n}\left(\mathbb{K}\right)\right)$. And $M_{m\times m}\left(R\right)$ is
isomorphic to $M_{m\times m}\left(\overline{R}\right)$ under the action of $G$, and for $\forall A\in M_{m\times m}\left(R\right)$,
$rank_{\mathbb{K}}\left( A\right)=rank_{\mathbb{K}}\left(G\left(A\right)\right)$.

Return to $M_{m\times m}(R)$. According to corollary 2, suppose $A\in M_{m\times m}(R)$, $G(A)\in M_{m\times m}(\overline{R})$. Thus, every
entry of $G(A)$ must have a diagonal shape like:
$$
[G(A)]_{i,j}^{\overline{R}} =
\begin{pmatrix}
S_{1}^{i,j}&&&\\
&S_{2}^{i,j}&&\\
&&\ddots &\\
&&&S_{d}^{ij}
\end{pmatrix}.
$$
Further, if
$$[A]_{i,j}^{R} = f_{i,j}(S) = \sum_{t=0}^{l-1}a_{t}S^{t},$$
then
$$[G(A)]_{i,j}^{\overline{R}} = g([A]_{i,j}^{R}) = P\cdot f_{i,j}(S)\cdot P^{-1} = \sum_{t=0}^{l-1}a_{t}(PSP^{-1})^{t}.$$
It's trivial that $S_{k}^{i,j} = f_{i,j}(S_{k})$ for all $i,j = 1,2,\cdots ,m$; $k=1,2,\cdots,d$.

Thus, by some row operations and column operations, we can transform $G(A)$ into a quasi-diagonal matrix over $M_{mn\times mn}(\mathbb{K})$:
$$
\Gamma_{0}\cdot G(A)\cdot \Gamma_{1} = \begin{pmatrix}
A_{1}&&&\\
&A_{2}&&\\
&&\ddots&\\
&&&A_{d}
\end{pmatrix}.
$$
$\Gamma_0$ and $\Gamma_1$ are products of some elementary matrix over $M_{mn\times mn}(\mathbb{K})$. $A_{t}\in M_{mn_{t}\times
mn_{t}}(\mathbb{K})$ was constructed by $S_{t}^{i,j}$ as follow:
$$
A_{t} =
\begin{pmatrix}
S_{t}^{1,1}&S_{t}^{1,2}&\cdots&S_{t}^{1,m}\\
S_{t}^{2,1}&S_{t}^{2,2}&&\vdots\\
\vdots&&\ddots&\\
S_{t}^{m,1}&\cdots&&S_{t}^{mm}
\end{pmatrix}.
$$
So $A_{t} \in M_{m\times m}(<S_{t}>)$, and we arrive at the conclusion that:
$$
rank_{\mathbb{K}}(A) = rank_{\mathbb{K}}(G(A)) = \sum_{t=1}^{d}rank_{\mathbb{K}}(A_{t}).
$$

\subsection{Linear complexity estimation model}
Let $R=<D> \in M_{n\times n}$, $D$ is a primitive circulant matrix with dimension $n$.

Obviously, the minimal polynomials of $D$ is $f(x) = x^{n} + 1$. Assume $f(x)$ have unique factorization $f(x) = \prod_{i=1}^{d}
p_{i}^{2^{\sigma}}(x)$, where $n/2^{\sigma}$ is exactly an odd integer.

From the conclusion of subsection 3.1, the linear complexity of the clock-controlled sequence $L(C) = rank_{\mathbb{K}}(\overline{C})$. At the
same time, $\overline{C} \in M_{m\times m}(R)$. Combining the matrix decomposition conclusions of subsection 3.2, we know that
$$
rank_{\mathbb{K}}(\overline{C}) = rank_{\mathbb{K}}(G(\overline{C})) = \sum_{t=1}^{d}rank_{\mathbb{K}}(\overline{C}_{t})
$$

At the same time, $\overline{C}_{t}$ is very similar to $\overline{C}$ and has the following form:
\begin{equation*}
\overline{C}_{t} =
\begin{pmatrix}
\hat{C}_{t} & O & \cdots & O\\
O & \hat{C}_{t} & & O\\
\vdots & &\ddots & \\
O & O & \cdots & \hat{C}_{t}\\
\end{pmatrix}
\cdot
\begin{pmatrix}
T_{t}^{s_{1}} & T_{t}^{s_{2}} & T_{t}^{s_{3}} &  & T_{t}^{s_{m} }\\
T_{t}^{s_{m} }D_{t}^{-1} & T_{t}^{s_{1}} & T_{t}^{s_{2} } & \cdots  &T_{t}^{s_{m-1}}\\
T_{t}^{s_{m-1} }D_{t}^{-1} & T_{t}^{s_{m} }D_{t}^{-1} & T_{t}^{s_{1}} & & T_{t}^{s_{m-2} }\\
&\vdots& &\ddots & \vdots\\
T_{t}^{s_{2}}D_{t}^{-1} & T_{t}^{s_{3} }D_{t}^{-1} & T_{t}^{s_{4} }D_{t}^{-1} &\cdots & T_{t}^{s_{1}}\\
\end{pmatrix}
\end{equation*}

Where
$$
P\cdot D\cdot P^{-1} = \begin{pmatrix}
D_{1}&&&\\
&D_{2}&&\\
&&\ddots&\\
&&&D_{d}
\end{pmatrix}.
$$
and $T_{t} = D_{t}^{v}$, where $v\times s_{m} \equiv 1 \mod n$.

$D_{t}$'s minimal polynomial is $p_{t}^{2^{\sigma}}(x)$, $D_{t}$ generate a commutative subalgebra, denote it by $R_{t} = <D_{t}>$. Recall the
theory of block-matrix, we know $R_{t} \cong \mathbb{F}_{2}[x]/(p_{t}^{2^{\sigma}}(x)) \triangleq H_{t}$. Set up $\phi_{t}$ to be the
isomorphism function from $R_{t}$ to $\mathbb{F}_{t}[x]/(p_{t}^{2^{\sigma}}(x))$, denote $\phi_{t}(D_{t})$ by $\alpha_{t}$, denote
$\phi_{t}(T_{t})$ by $\beta_{t}$. Thus, $\beta_{t} = \alpha^{v} \mod p_{t}^{2^{\sigma}}(x)$. Furthermore, consider the projection $\delta_{t}$
from $\mathbb{F}_{2}[x]/(p_{t}^{2^{\sigma}}(x))$ to field $\mathbb{F}_{2}[x]/(p_{t}(x))\triangleq \overline{H_{t}}$:
$$
\delta_{t}(\overline{h(x)}) = \overline{(h(x)\mod p_{t}(x))}
$$

Let $\delta_{t}(\alpha_{t}) = \overline{\alpha_{t}}$, $\delta_{t}(\beta_{t}) = \overline{\beta_{t}}$.

Denoted matrix $M_{t} \in M_{m\times m}(R_{t})$ and $\overline{M_{t}} \in M_{m\times m}(\overline{H_{t}})$:

\begin{equation*}
M_{t}^{R_{t}} =
\begin{pmatrix}
T_{t}^{s_{1}} & T_{t}^{s_{2}} & T_{t}^{s_{3}} &  & T_{t}^{s_{m} }\\
T_{t}^{s_{m} }D_{t}^{-1} & T_{t}^{s_{1}} & T_{t}^{s_{2} } & \cdots  &T_{t}^{s_{m-1}}\\
T_{t}^{s_{m-1} }D_{t}^{-1} & T_{t}^{s_{m} }D_{t}^{-1} & T_{t}^{s_{1}} & & T_{t}^{s_{m-2} }\\
&\vdots& &\ddots & \vdots\\
T_{t}^{s_{2}}D_{t}^{-1} & T_{t}^{s_{3} }D_{t}^{-1} & T_{t}^{s_{4} }D_{t}^{-1} &\cdots & T_{t}^{s_{1}}\\
\end{pmatrix}
\end{equation*}

\begin{equation*}
\overline{M_{t}} =
\begin{pmatrix}
\overline{\beta_{t}}^{s_{1}} & \overline{\beta_{t}}^{s_{2}} & \overline{\beta_{t}}^{s_{3}} &  & \overline{\beta_{t}}^{s_{m} }\\
\overline{\beta_{t}}^{s_{m} }\overline{\alpha_{t}}^{-1} & \overline{\beta_{t}}^{s_{1}} & \overline{\beta_{t}}^{s_{2} } & \cdots
&\overline{\beta_{t}}^{s_{m-1}}\\
\overline{\beta_{t}}^{s_{m-1} }\overline{\alpha_{t}}^{-1} & \overline{\beta_{t}}^{s_{m} }\overline{\alpha_{t}}^{-1} &
\overline{\beta_{t}}^{s_{1}} & & \overline{\beta_{t}}^{s_{m-2} }\\
&\vdots& &\ddots & \vdots\\
\overline{\beta_{t}}^{s_{2}}\overline{\alpha_{t}}^{-1} & \overline{\beta_{t}}^{s_{3} }\overline{\alpha_{t}}^{-1} & \overline{\beta_{t}}^{s_{4}
}\overline{\alpha_{t}}^{-1} &\cdots & \overline{\beta_{t}}^{s_{1}}\\
\end{pmatrix}
\end{equation*}

Since Theorem 4,
$$
rank_{\mathbb{K}}(\overline{C}_{t}) \geq rank_{\mathbb{K}}(\hat{C}_{t})\times rank_{R_{t}}(M_{t}^{R_{t}}).
$$

Since Theorem 5,
$$
rank_{\mathbb{K}}(\overline{C}_{t}) \geq rank_{\mathbb{K}}(\hat{C}_{t})\times rank_{R_{t}}(M_{t}^{R_{t}}) =
rank_{\mathbb{K}}(\hat{C}_{t})\times rank_{\overline{H_{t}}}(\overline{M_{t}}).
$$

Finally, we get a Linear complexity lower bound estimation inequality.
\begin{theorem}
$$
L(C) = \sum_{t=1}^{d}rank_{\mathbb{K}}(\overline{C}_{t}) \geq \sum_{t=1}^{d}rank_{\mathbb{K}}(\hat{C}_{t})\times
rank_{\overline{H_{t}}}(\overline{M_{t}})
$$
\end{theorem}

%The final problem translates into how to estimate $rank_{\mathbb{K}}(\hat{C}_{t})$ and $rank_{\overline{H_{t}}}(\overline{M_{t}})$.%

The last problem turns into how to estimate $rank_{\mathbb{K}}(\hat{C}_{t})$ and $rank_{\overline{H_{t}}}(\overline{M_{t}})$.

\subsubsection{Estimate  $rank(\hat{C}_{t})$ over $F_{2}$:}

Since $C^1=\left\{c_{1+m\cdot i}\right\}_{i=0}^\infty$, so $\hat{C}=M_{cir}\left(C^1\right)\cdot T^{-s_1}=\left(\sum_{i=0}^{n-1}{c_{1+m\cdot
i}D^i}\right)\cdot D^{-s_1v}$. Thus:
$$
\hat{C}_{t}=\left(\sum_{i=0}^{n-1}{c_{1+m\cdot i}D_t^i}\right)\cdot D_t^{-s_1v}
$$

That means, $rank_{F_{2}}(\hat{C}_{t})$ equals to rank of matrix $\sum_{i=0}^{n-1}{c_{1+m\cdot i}D_t^i}$.

Let the formal power series $F^1\left(x\right)=\sum_{i=0}^{\infty}c_{1+m\cdot i}x^i$, be the generating function of the sequence $C^1$, and let
$H\left(x\right)\in F_{2}\left[x\right]$, be the minimum generator Polynomial of $C^1 $. The order of $H\left(x\right)$ is equal to the linear
complexity of $C^1$. There exists polynomial $P\left(x\right)\in F_{2}\left[x\right]$ with number less than $l$ such that the following
constant equation holds, $H\left(x\right)$ and $P\left(x\right)$ are coprime.\cite{DenguoF1999}
$$F^1\left(x\right)=\frac{P\left(x\right)}{H\left(x\right)}$$

Let $\overline{F^1}\left(x\right)=\sum_{i=0}^{n-1}c_{1+m\cdot i}x^i$, then the power series of the form
$$F^1\left(x\right)=\frac{\overline{F^1}\left(x\right)}{1+x^n}$$
and $H(x)| 1+x^{n}$. This is a conclusion that comes from the minimal property of $H\left(x\right)$.

Thus,
$$\overline{F^1}\left(x\right)\cdot H\left(x\right)=P\left(x\right)\cdot\left(1+x^n\right)$$
The equation no longer needs to be discussed under the formal power series sense and goes back to the polynomial ring $F_{2}\left[x\right]$.

As we know, $x^n+1=\prod_{i=1}^{d}{p_i^{2^\delta}\left(x\right)}$ and $H(x)|(1+x^n)$. Assume
$H\left(x\right)=\prod_{i=1}^{d}{p_i^{h_i}\left(x\right)}$, where $0\leq h_{i} \leq 2^{\sigma}$. Then:
$$\overline{F^1}\left(x\right)=P\left(x\right)\cdot\prod_{i=1}^{d}{p_i^{2^\delta-h_i}\left(x\right)}$$

\begin{itemize}
    \item When $h_{t} = 0$, Since the characteristic(minimal) polynomial of the matrix $D_t$ is $p_t^{2^\delta}\left(x\right)$,
        $p_t^{2^\delta}\left(D_t\right)$ is a zero square matrix. Led to
    $$\overline{F^1}\left(D_t\right)=P\left(D_t\right)\cdot\prod_{i\neq t}{p_i^{2^\delta-h_i}\left(D_i\right)}\cdot
    p_t^{2^\delta}\left(D_t\right)=O$$
    $$rank\left(\hat{C}_{t}\right)=rank(\overline{F^1}\left(D_t\right)) = 0$$
    \item When $h_{t} > 0$, Since $H\left(x\right)$ is coprime to $P\left(x\right)$, $p_t\left(x\right)$ is coprime to $P\left(x\right)$.
        Therefore, $P\left(D_t\right)$ is still a full-rank square (because the root sets of $P\left(x\right)=0$ does not include any
        characteristic root of $D_t$). Thus,
        $$rank\left(\hat{C}_{t}\right)=rank\left(\overline{F^1}\left(D_t\right)\right)=rank\left(p_t^{2^\delta-h_t}\left(D_t\right)\right)\geq
        h_t\times deg\left(p_t\left(x\right)\right)$$
\end{itemize}

Combining these two cases, the following inequalities can be derived.
$$rank\left(\hat{C}\right)=\sum_{t=1}^{d}{rank\left(\hat{C}_t\right)}\geq\sum_{t=1}^{d}\left(h_t\times d e
g\left(p_t\left(x\right)\right)\right)=deg\left(H\left(x\right)\right)$$

Notice that both left and right of the inequality are equal to the $L\left(C^1\right)$, so the inequality equal sign holds constant. That is,
for $\forall 1\leq t\leq d$, we have $$rank\left(\hat{C}_t\right)=h_t\times deg\left(p_t\left(x\right)\right).$$

\subsubsection{Estimate $rank_{\overline{H_{t}}}(\overline{M_{t}})$:}

It's easy to see that $\overline{M_{t}}$ is an $\overline{\alpha_{t}}^{-1}$-circulant matrix over field $\overline{H_{t}}$. Let $E_{t}(x) =
\sum_{i=0}^{m-1}\overline{\beta_{t}}^{s_{i+1}}x^{i} \in \overline{H_{t}}[x]$,
$J_{t}(x)=x^m+\overline{\alpha_{t}}^{-1}\in\overline{H_t}\left[x\right]$. Use theorem 3, denote The degree of the greatest common factor of
$E_{t}(x)$ and $J_{t}(x)$ by $g_{t}$, then
$$rank_{\overline{H_{t}}}(\overline{M_{t}}) = m - g_{t}.$$

In summary, we get a final inequality of rank:
$$rank\left(M_{t}\right)\geq rank\left(\hat{C}_{t}\right)\cdot rank_{R_t}\left({\overline{M_t}}^{R_t}\right)=h_t\times
deg\left(p_t\left(x\right)\right)\times\left(m-g_t\right)$$

After accumulation:
\begin{theorem}
$$L(C) \geq \sum_{t=1}^{d}h_t\times deg\left(p_t\left(x\right)\right)\times\left(m-g_t\right)$$
\end{theorem}

\subsection{Section summary}
This section analyzed the lower bound on the linear complexity of the basic feedforward clock-controlled sequence.

The first step is to correspond the linear complexity to the rank of the cyclic matrix. After that, the matrix is organized according to a
particular sampling law. In this way, the matrix becomes a matrix on a circulant matrix ring.

However, the matrix on a normal commutative ring is not easy to count the rank. So further quasi-diagonalization is performed for each matric
block at the same time. Then the goal becomes to compute the sum of the ranks of all matrices $M_{t}$ on the diagonal.

Using Theorem 4 and Theorem 5, we can successfully estimate the rank of the matrix $M_{t}$.

Through such a series of transformations,  we decompose the problem to each subfield. In this way, the enormous problem of overall linear
complexity becomes a collection of several minor problems. Finally, we obtained a valuable conclusion.

The following section gives a new stream cipher LIFI-128 using a nonlinear drive module reference to the LILI-128 algorithm. This kind of
stream cipher's complexity is impossible to be estimated by traditional results. However, our new method can solve its linear complexity
problem.
\section{LIFI-128, and it's linear complexity}
\subsection{Description of LIFI-128}
We give an example that was set up to follow the LILI-128 algorithm. The clock-control subsystem uses a pseudorandom binary sequence produced
by a regularly clocked LFSR, $LFSR_{a}$, of length $39$, and a function, $f_{a}$, operating on some contents of $LFSR_{a}$ to produce a
pseudorandom integer sequence, $A = \{a_{i}\}_{\infty}$, and $a_{i} \in \{1,2,3,4\}$. The feedback polynomial of $LFSR_{a}$ is chosen to be a
primitive polynomial. Moreover, the initial state of $LFSR_{a}$ must not be all zero. It follows that $LFSR_{a}$ produces a maximum-length
sequence of period $m = 2^{39} - 1$. Set $f_{a}$ to be boolean balance function every bit, so $s_{m} = 2^{39-1}(2^2+1)-1$.

The data-generation subsystem uses the integer sequence $A$ to control the clocking of a binary FCSR\cite{klapper19932}, $FCSR_{b}$, of length
$89$.  The lowest content bit of $FCSR_{b}$ will generate a binary sequence $B = \{b_{i}\}_{\infty}$. The feedback integer of $FCSR_{b}$ is
chosen to be a safe prime integer $q = 2p + 1$, where $p$ is a prime, $\lfloor\log_{2}(q)\rfloor = 89$. And the initial state of $FCSR_{b}$ is
not zero, $2$ is a primitive elements of $\mathbb{F}_{q}^{*}$. It follows that $FCSR_{b}$ produces a nonlinear binary sequence with period $n =
2p$, linear complexity $L(B) = p+1$. Specially, in this example, we assume $2$ also is a primitive elements of $\mathbb{F}_{p}^{*}$, and
$gcd(p-1,39)=1$.

Notice that $n = 2p$ is coprime with $s_{m} = 2^{39-1}(2^2+1)-1$. Define the clock-controlled sequence $C = \{c_{i} = b_{s_{i}}\}_{\infty}$ as
the keystream, thus $C$ get a maximum-length of period $mn = (2^{39}-1)\times (2p) \approx 2^{128}$.

\begin{figure}
\begin{center}
\includegraphics[height=4cm]{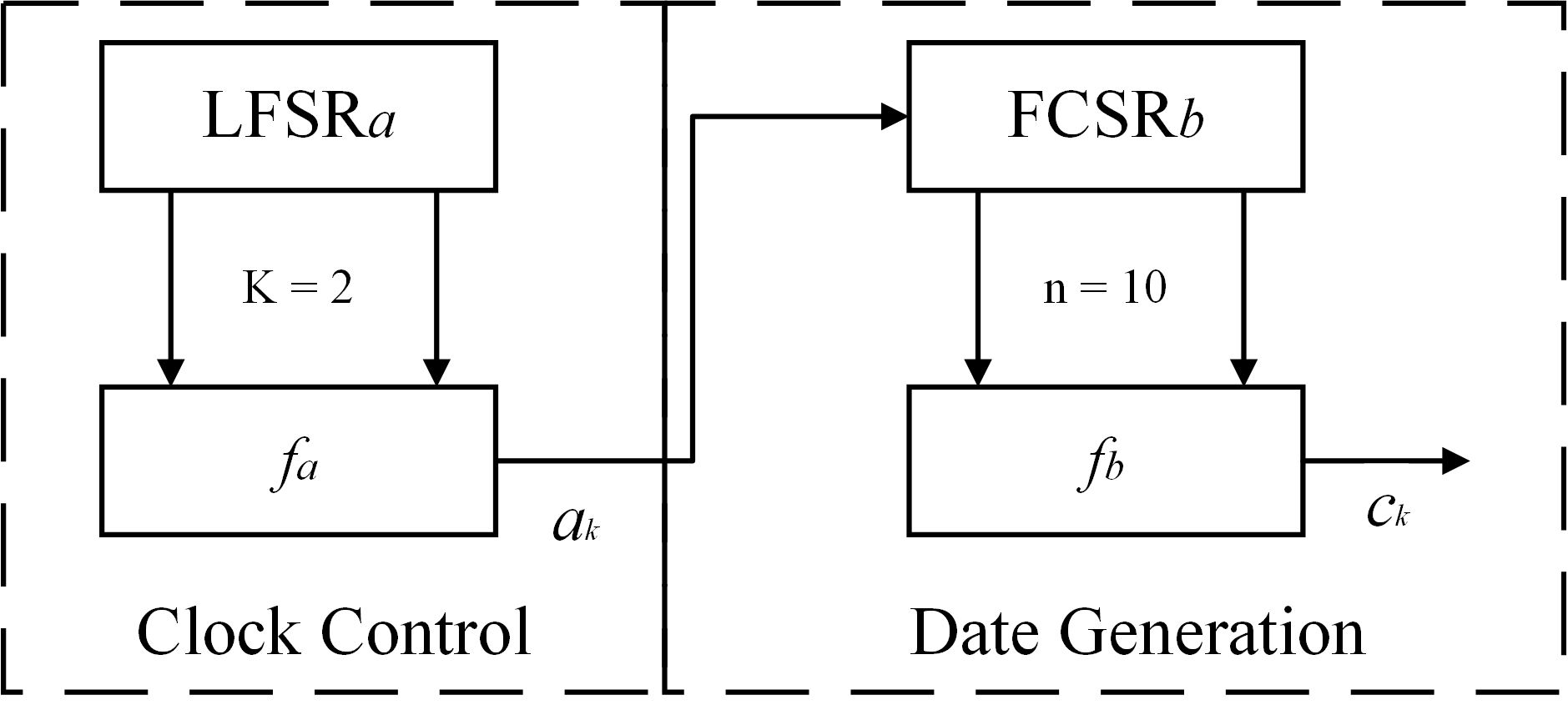}
\caption{The Structure of LIFI-128}
\end{center}
\end{figure}

\subsection{Linear Complexity}
We are going to prove that the linear complexity of $C$ has a lower bound $(L(B)-2)m + 2= (p-1)(2^{39}-1) + 2$.

Consider the $n\times n$ circulant matrices ring $R$ which was generated by $D$. It's easy to show that $R$ is a commutative algebra over
$\mathbb{F}_{2}$, thus, for any period $n$ sequence $S$, $M_{cir}(S) \in R$. As this result, the clock-matrix is a $m\times m$ matrix over $R$.

It is obvious that the minimal polynomial of $D$ equals to $f(x) = x^{n} + 1$. $f(x)$ has a decomposition over $\mathbb{F}_{2}$:
$$f(x) = x^{2p} + 1 = (x^{p} + 1)^{2} = (x+1)^{2}(\sum_{i=0}^{p-1}x^i)^{2}.$$
As $2$ is a primitive elements of $\mathbb{F}_{p}^{*}$, $\sum_{i=0}^{p-1}x^{i}$ is reduced in $\mathbb{F}_{2}[x]$. Assume $f(x) =
p_{1}^{2}(x)p_{2}^{2}(x)$, where $p_{1}(x) = x + 1$, $p_{2}(x) = \sum_{i=0}^{p-1}x^{i}.$ Thus, we can find a $n\times n$ nonsingular matrix $P$
over $\mathbb{F}_{2}$ such that $P\cdot S \cdot P^{-1}$ is a quasi-diagonalization on $S\in R$.
$$
P\cdot S \cdot P^{-1} =
\begin{pmatrix}
Q_{1}&O\\
O&Q_{2}
\end{pmatrix}
$$
In above formula $Q_{1}$ is the factor relates to $p_{1}^{2}(x)$, $Q_{2}$ is factor relates to $p_{2}^{2}(x)$. As $degree(p_{1}) = 1,
degree(p_{2}) = p-1$, $Q_{1}\in M_{2\times 2}(\mathbb{F}_{2})$, $Q_{2} \in M_{2(p-1)\times 2(p-1)}(\mathbb{F}_{2})$.

Notice that $\hat{C},T,D\in R$, we denote:
$$
P\hat{C}P^{-1} =
\begin{pmatrix}
\hat{C}_{1}&O\\
O&\hat{C}_{2}
\end{pmatrix},\ \
PTP^{-1} =
\begin{pmatrix}
T_{1}&O\\
O&T_{2}
\end{pmatrix},\ \
PDP^{-1} =
\begin{pmatrix}
D_{1}&O\\
O&D_{2}
\end{pmatrix}
$$

By theory of FCSR\cite{klapper19932}, the minimial generator Polynomial of sequence $B'=\{b_{s_{m}i+1}\}_{i=0}^{\infty}$ is $H(x) =(x+1)(x^p+1)
= p_{1}^{2}(x) p_{2}(x)$. Thus $rank(\hat{C}_{1}) = 2$, $rank(\hat{C}_{2}) = p-1$.

In field $F_{2}[x]/(x+1) = F_{2}$, $\overline{\alpha_{1}} = \overline{\beta_{1}} = \overline{1}$. Then:
\begin{equation*}
\overline{M_{1}} =
\begin{pmatrix}
\overline{\beta_{1}}^{s_{1}} & \overline{\beta_{1}}^{s_{2}} & \overline{\beta_{1}}^{s_{3}} &  & \overline{\beta_{1}}^{s_{m} }\\
\overline{\beta_{1}}^{s_{m} }\overline{\alpha_{1}}^{-1} & \overline{\beta_{1}}^{s_{1}} & \overline{\beta_{1}}^{s_{2} } & \cdots
&\overline{\beta_{1}}^{s_{m-1}}\\
\overline{\beta_{1}}^{s_{m-1} }\overline{\alpha_{1}}^{-1} & \overline{\beta_{1}}^{s_{m} }\overline{\alpha_{1}}^{-1} &
\overline{\beta_{1}}^{s_{1}} & & \overline{\beta_{1}}^{s_{m-2} }\\
&\vdots& &\ddots & \vdots\\
\overline{\beta_{1}}^{s_{2}}\overline{\alpha_{1}}^{-1} & \overline{\beta_{1}}^{s_{3} }\overline{\alpha_{1}}^{-1} & \overline{\beta_{1}}^{s_{4}
}\overline{\alpha_{1}}^{-1} &\cdots & \overline{\beta_{1}}^{s_{1}}\\
\end{pmatrix}
=
\begin{pmatrix}
\overline{1}&\overline{1}&\cdots&\overline{1}\\
\overline{1}&\overline{1}&\cdots&\overline{1}\\
\vdots & \vdots & \ddots & \vdots \\
\overline{1}&\overline{1}&\cdots&\overline{1}\\
\end{pmatrix}.
\end{equation*}

It's easy to see $rank(\overline{M_{1}}) = 1$.

Next, we are going to count follow matrix $\overline{M_{2}}$.
\begin{equation*}
\overline{M_{2}} =
\begin{pmatrix}
\overline{\beta_{2}}^{s_{1}} & \overline{\beta_{2}}^{s_{2}} & \overline{\beta_{2}}^{s_{3}} &  & \overline{\beta_{2}}^{s_{m} }\\
\overline{\beta_{2}}^{s_{m} }\overline{\alpha_{2}}^{-1} & \overline{\beta_{2}}^{s_{1}} & \overline{\beta_{2}}^{s_{2} } & \cdots
&\overline{\beta_{2}}^{s_{m-1}}\\
\overline{\beta_{2}}^{s_{m-1} }\overline{\alpha_{2}}^{-1} & \overline{\beta_{2}}^{s_{m} }\overline{\alpha_{2}}^{-1} &
\overline{\beta_{2}}^{s_{1}} & & \overline{\beta_{2}}^{s_{m-2} }\\
&\vdots& &\ddots & \vdots\\
\overline{\beta_{2}}^{s_{2}}\overline{\alpha_{2}}^{-1} & \overline{\beta_{2}}^{s_{3} }\overline{\alpha_{2}}^{-1} & \overline{\beta_{2}}^{s_{4}
}\overline{\alpha_{2}}^{-1} &\cdots & \overline{\beta_{2}}^{s_{1}}\\
\end{pmatrix}
\end{equation*}

$\overline{\beta_{2}} = \overline{\alpha_{2}}^{v}$, where $v\times s_{m} \equiv 1 \mod n$.

Notice that $\overline{M_{2}}$ is a $m\times m$ matrix over field $\mathbb{F}_{2}[x]/(p_{2}(x))$, in particular, $\overline{M_{2}}$ is a
$\overline{\alpha_{2}}^{-1}$-circulant matrix.

With the help of Theorem 3, we know $\overline{M_{2}}$ is non-singular if and only if $w(x) =
\sum_{i=1}^{m}\overline{\beta_{2}}^{s_{i}}x^{i-1}$ doesn't have common root with $x^{m} - \overline{\alpha_{2}}^{-1} = 0$, or
$\overline{M_{2}}$ has a zero eigenvalue.

As $m$ is coprime with the order $p$ of $\overline{\alpha_{2}}^{-1}$, there is a integer $k$ such that $(\overline{\alpha_{2}}^{k})^{m} =
\overline{\alpha_{2}}^{-1}$. Assume $\{\xi^{i}|i=0,1,\cdots,m-1\}$ is all roots of equation $x^{m} - 1 = 0$, then roots set of $x^{m} -
\overline{\alpha_{2}}^{-1} = 0$ is $\{\xi^{i}\overline{\alpha_{2}}^{k}|i=0,1,\cdots,m-1\}$. If $\xi^{j}\overline{\alpha_{2}}^{k}$ is a root of
$w(x)$, which means:
$$
\begin{aligned}
w(\xi^{j}\overline{\alpha_{2}}^{k}) =& \sum_{i=1}^{m}\overline{\beta_{2}}^{s_{i}}(\xi^{j}\overline{\alpha_{2}}^{k})^{i-1}\\
 =& \sum_{i=1}^{m}\overline{\alpha_{2}}^{s_{i}v + (i-1)k}(\xi^{j})^{i-1}
\end{aligned}
$$
Define another function $g(x) = \sum_{i=1}^{m}\overline{\alpha_{2}}^{s_{i}v + (i-1)k}x^{i-1}$, it's easy to see that\\ $gcd(w(x),x^{m} -
\overline{\alpha_{2}}^{-1}) \neq 1$ equals to $gcd(g(x), x^{m} - 1) \neq 1$.

The proof is by contradiction, suppose that there is $0\leq i' \leq m-1$, $\xi' = \xi^{i'}$ is a common root of $g(x)$ and $x^{m}-1$. Notice
that $\xi'^{m+1} = \xi'^{2^{39}} = \xi'$, we calculate the $2^{39}$th power of $g(\xi')$. It shows:
$$
\begin{aligned}
0 =& (g(\xi'))^{2^{39}}\\
=& \sum_{i=1}^{m}\overline{\alpha_{2}}^{(s_{i}v + (i-1)k)\times 2^{39}}(\xi'^{2^{39}})^{i-1}\\
=& \sum_{i=1}^{m}(\overline{\alpha_{2}}^{2^{39}})^{s_{i}v+(i-1)k}\xi'^{i-1}
\end{aligned}
$$

Define function
$$
\begin{aligned}
h(\overline{\alpha_{2}}) =& \sum_{i=1}^{m}\xi'^{i-1}(\overline{\alpha_{2}})^{(s_{i}v+(i-1)k \mod p)}\\
 =& \sum_{i=0}^{p-1}[\sum_{s_{j}v+(j-1)k \equiv i}\xi'^{j}]\overline{\alpha_{2}}^{i}
\end{aligned}
$$

Above formula shows that if $\overline{\alpha_{2}}$ is a root of $h(\overline{\alpha_{2}}) = 0$, then $\overline{\alpha_{2}}^{2^{39}}$ will
also be a root of $h(\overline{\alpha_{2}})=0$. Now we get a set $\Omega = \{\overline{\alpha_{2}}^{2^{39i}}| i = 0, 1, \cdots\}$, any elements
of this set would be a root of $h(\overline{\alpha_{2}}) = 0$. Recall that $\overline{\alpha_{2}}$'s order is $p$, and $2$ is a primitive
element of $\mathbb{F}_{p}^{*}$, so $\overline{\alpha_{2}}^{2^{39a}} = \overline{\alpha_{2}}^{2^{39b}}$ if and only if $39a \equiv 39b \mod
(p-1)$. Thus $\#|\Omega| = p-1$. As degree of $h(\overline{\alpha_{2}})$ less than or equal to $p-1$, $\Omega$ must be all roots set of
$h(\overline{\alpha_{2}}) = 0$. Thus, $h(\overline{\alpha_{2}}) = \xi^{*}\prod_{i=0}^{p-1}(\overline{\alpha_{2}} -
\overline{\alpha_{2}}^{2^{39i}})$, $\xi^{*}$ is a constant.

Denote $\sum_{s_{j}v+(j-1)k \equiv i}\xi'^{j}$ by $\varepsilon_{i}$, $h(\overline{\alpha_{2}}) =
\sum_{i=0}^{p-1}\varepsilon_{i}\overline{\alpha_{2}}^{i}$ and $\xi^{*} = \varepsilon_{p-1} \neq 0$. Notice $\Omega \subseteq
\mathbb{F}_{2}[x]/p_{2}(x)$, suppose that $\varepsilon_{i_{0}} \neq 0$, then $\varepsilon_{i_{0}}/\xi^{*}$ must be an element in field
$\mathbb{F}_{2}[x]/p_{2}(x)$, as it is represented as an element generated by $\Omega$ over field $\mathbb{F}_{2}[x]/p_{2}(x)$. Denote this
field by $\mathbb{F}_{2^{p-1}}$.

In the same way,
$$\varepsilon_{i_{0}}/\xi^{*} = [\sum_{s_{j}v+(j-1)k \equiv i_{0}}\xi'^{j}]/[\sum_{s_{j}v+(j-1)k \equiv p-1}\xi'^{j}].$$

In fact, we can show that $\#\{1\leq j \leq m| s_{j}v+(j-1)k \equiv i_{0} \mod p\} \leq 1$ for any $i_{0}$. Suppose there are $1 \leq j < j'
\leq m$ such that $s_{j}v+(j-1)k \equiv s_{j'}v+(j'-1)k \mod p$. Thus, $s_{m}m(s_{j'} - s_{j})v + s_{m}m(j' - j)k \equiv 0 \mod p$, equals to
$(s_{j'}-s_{j})m - s_{m}(j' - j) \equiv 0 \mod p$. Notice that
$$
\begin{aligned}
|(s_{j'}-s_{j})m - s_{m}(j' - j)| \leq & |(s_{j'}-s_{j})m| + |s_{m}(j' - j)|\\
\leq & 4(j'-j)m+s_{m}(j'-j)\\
=& (5\times 2^{38}+3)(j'-j)\\
\leq & 2^{41}\times 2^{39}\\
= & 2^{80} < p.
\end{aligned}
$$
Thus $(s_{j'}-s_{j})m = s_{m}(j' - j)$. Because $gcd(m,s_{m}) = gcd(2^{39}-1, 5\times 2^{38}-1) = 1$, we get $m|(j'-j)$, conflicts with $1\leq
j'-j \leq m-1$ and implies $\#\{1\leq j \leq m| s_{j}v+(j-1)k \equiv i_{0} \mod p\} \leq 1$.

This fact shows that $\varepsilon_{i_{0}}/\xi^{*} = 0$, or $\varepsilon_{i_{0}}/\xi^{*} = \xi'^{s}$ for an integer $s$. As
$\varepsilon_{i_{0}}/\xi^{*}\in \mathbb{F}_{2^{p-1}}$, $\xi'^{m} = 1$\ and\ $gcd(m,2^{p-1}-1) = 1$, those facts led to $\xi'^{s} = 1$. But it's
easy to confirm that $g(\xi') = g(1) \neq 0$.This result conflicts with $g(\xi') = 0$. Based on these facts, $g(x)$ doesn't have common root
with $x^{m}-1$, $\overline{M_{2}}$ is non-singular, $rank(\overline{M_{2}}) = m$

Thus, recall theorem 7,
$$
\begin{aligned}
L(C) &\geq \sum_{t=1}^{2}h_{t}\times deg(p_{t}(x))\times (m-g_{t})\\
 &= 2\times 1\times 1 + 1\times (p-1)\times m\\
 & = m(p-1) + 2.\\
\end{aligned}
$$
$C$'s rank greater then $(p-1)m +2 = (p-1)(2^{39}-1) + 2$, we get a linear complexity lower bound of clock-controlled sequence
$\{c_{i}\}_{\infty}$.
\subsection{Section summary}
This section modifies the LILI-128 algorithm so that its controlled sequence becomes nonlinearly driven with extremely high linear complexity.
We call the new algorithm LIFI-128.

None of the published linear complexity analysis methods give a good result for LIFI-128. However, our new model can solve this type of problem
very well. The practical value of the linear complexity lower bound estimation method proposed in this paper is fully illustrated.
\section{Conclusion}
The feedforward clock control structure is a hardware-friendly and widely used structure for designing sequence encryption algorithms. Its
basic structure is that two sequence generators connect in series. The first generator is regular output and the second generator
clock-controlled by the output of the first generator.

In this paper, we research the feedforward clock-controlled sequence structure by new methods such as circulant matrix and matrix over the
ring. Finally, the resulting complexity estimation inequalities can widely apply to the analysis of cryptographic properties of the
clock-controlled structure.  The traditional result base on cyclotomic polynomials over finite fields is not practical when the controlled
sequence is nonlinear. The results presented in this paper can be exactly effective for the analysis of clock-controlled cryptographic systems
whether the drive module is linear or nonlinear.

\subsection*{Acknowledgements.}
We thank the anonymous reviewers for their helpful comments. This work was supported by the National Natural Science Foundation of China(Grant
No.6207211, No.61672059) and the National Key R$\&$D Program of China 2017YFB0802000.

\bibliographystyle{splncs04}
\bibliography{ins21}
\end{document}